# Status and Challenges for FCC-ee

Michael Benedikt, Katsunobu Oide*, Frank Zimmermann
Mail to: michael.benedikt@cern.ch , katsunobu.oide@cern.ch ,
frank.zimmermann@cern.ch
CERN, Route de Meyrin, 1211 Geneva 23, Switzerland
* also KEK, Tsukuba, Japan

Anton Bogomyagkov, Eugene Levichev
Mail to: a.v.bogomyagkov@inp.nsk.su , e.b.levichev@inp.nsk.su
BINP, Novosibirsk, Russia

Mauro Migliorati
Mail to: mauro.migliorati@uniroma1.it
Sapienza University, Roma, Italy

Uli Wienands
Mail to: uli@slac.stanford.edu
SLAC National Accelerator Laboratory, Menlo Park, U.S.A.

*Abstract*

We report the design status and beam dynamics challenges for the electron-positron branch of the Future Circular Collider (FCC) study, as of August 2015. After recalling motivation and physics requirements for the FCC-ee, we briefly discuss configurations and parameters, collider layout, the superconducting RF system, possible staging scenarios, final-focus optics, interaction-region (IR) issues, machine detector interface and IR synchrotron radiation, dynamic aperture, beam-beam effects, top-up injection, mono-chromatization option, impedances, instabilities, energy calibration and polarization, and SuperKEKB as a key demonstrator, before presenting conclusions and outlook.

**Motivation**

In response to a request from the 2013 Update of the European Strategy for Particle Physics [1], the global Future Circular Collider (FCC) study is designing a 100-TeV proton collider (FCC-hh) in a new ~100 km tunnel near Geneva, a high-luminosity electron-positron collider (FCC-ee) as a potential intermediate step, and a lepton-hadron option (FCC-he). The FCC study comprises accelerators, technology, infrastructure, detector, physics, concepts for worldwide data services, international governance models, and implementation scenarios.

The specific goals of the FCC-ee collider call for luminosities above $10^{36}$ cm$^{-2}$s$^{-1}$ per interaction point at the *Z* pole or some $10^{34}$ cm$^{-2}$s$^{-1}$ at the *ZH* production peak, and pushing the beam energy up to ≥175 GeV, with a total synchrotron-radiation power not exceeding 100 MW. The extremely high luminosity and resulting short beam lifetime (due to radiative Bhabha scattering) are sustained by top-up injection [2].

An FCC-ee parameter baseline as well as high-luminosity crab-waist options were described in [3] and [4], respectively.

**Physics Requirements**

The FCC-ee should achieve highest possible luminosities over a wide range of beam energies, from 35 GeV to ≈200 GeV, supporting extremely high precision tests of the standard model as well as unique searches for rare decays.

The FCC-ee physics programme [5] includes: (1) $\alpha_{QED}$ studies (with energies as low as 35 GeV) to measure the running coupling constant close to the Z pole; (2) operation on the Z pole (45.5 GeV), where FCC-ee would serve as a 'TeraZ' factory for high precision $M_Z$ & $\Gamma_Z$ measurements and allow searches for extremely rare decays (enabling the hunt for sterile right-handed neutrinos); (3) running at the *H* pole (63 GeV) for *H* production in the *s* channel, with mono-chromatization, e.g. to map the width of the Higgs; (4) operation at the W pair production threshold (~80 GeV) for high precision $M_W$ measurements; (5) operation in *ZH* production mode (maximum rate of *H*'s) at 120 GeV; (6) operation at and above the $t\bar{t}$ threshold (~175 GeV); and (7) operation at energies above 175 GeV per beam should a physics case for the latter be made.

Scaling from LEP and LEP2 some beam polarization is expected for beam energies up to ≥80 GeV [6], which will be exploited for precise energy calibration using resonant depolarization.

The collider may be optimized for operation at 120 GeV (Higgs factory), with 45.5 GeV (TeraZ factory) as second priority.

**Configurations and Parameters**

The FCC-ee layout must be compatible with the tunnel infrastructure for the hadron collider FCC-hh. Some of its key elements are: (a) a double ring with separate beam pipes, magnet-strength tapering (to compensate for the energy sawtooth due to synchrotron radiation), and independent optics control for the counter-circulating electron and positron beams, colliding at a total crossing angle of 30 mrad; (b) top-up injection based on a fast-cycling booster synchrotron housed in the same large tunnel with bypasses around the particle-physics detectors; and (c) local chromatic correction of the final-focus systems.

The range of FCC-ee beam parameters is indicated in Table 1, for simplicity showing numbers for (only) three different operation modes. The beam current varies greatly with beam energy, ranging from a few mA, as at LEP2, to 1.5 A, similar to the B factories. As a design choice, the total synchrotron radiation power has been limited to 100 MW, about 4 times the synchrotron-radiation power of LEP2. For a roughly four times larger machine this results in comparable radiation power per unit length. The present numbers might translate into a total wall plug power around 300 MW. The estimated luminosity numbers scale linearly with the synchrotron-radiation power. Other important choices to be made, or to be confirmed, are the number of collisions points (2 or 4), the crossing angle (30 mrad in total), and the collision scheme (crab waist?).

Presently there is a trend to transit from the original baseline [3], in which the arc cell length is varied so as to maintain almost constant geometrical emittance at all beam energies, to the crab-waist scheme, for which the smallest possible transverse emittances are desired at all energies. On the *Z* pole, the crab-waist approach [4] could achieve about ten times more luminosity than the baseline whereas at the high energy operation points the performance of the two optics variants is about equal. Figure 1 displays the expected luminosity per IP as a function of c.m. energy, assuming crab-waist collisions at two points.

**Table 1:** Key parameters for FCC-ee, at three beam energies, compared with LEP2. The parameter ranges indicated reflect a sensitivity to the number of IPs and to the choice of collision scheme ("baseline" [3] with varying arc cell length and small crossing angle, or a crab-waist scheme based on a larger crossing angle and constant cell length [4]).

| Parameter | FCC-ee | | | LEP2 |
|---|---|---|---|---|
| energy/beam [GeV] | 45 | 120 | 175 | 105 |
| bunches/beam | 13000- 60000 | 500- 1400 | 51- 98 | 4 |
| beam current [mA] | 1450 | 30 | 6.6 | 3 |
| luminosity/IP x $10^{34}$cm$^{-2}$s$^{-1}$ | 21 - 280 | 5 - 11 | 1.5 - 2.6 | 0.0012 |
| vertical IP $\beta^*$ [mm] | 1 | 1 | 1 | 50 |
| geom. hor. emittance [nm] | 0.1-30 | 1 | 2 | 22 |
| energy loss/turn [GeV] | 0.03 | 1.67 | 7.55 | 3.34 |
| synchrotron power [MW] | 100 | | | 22 |
| RF voltage [GV] | 0.2-2.5 | 3.6-5.5 | 11 | 3.5 |

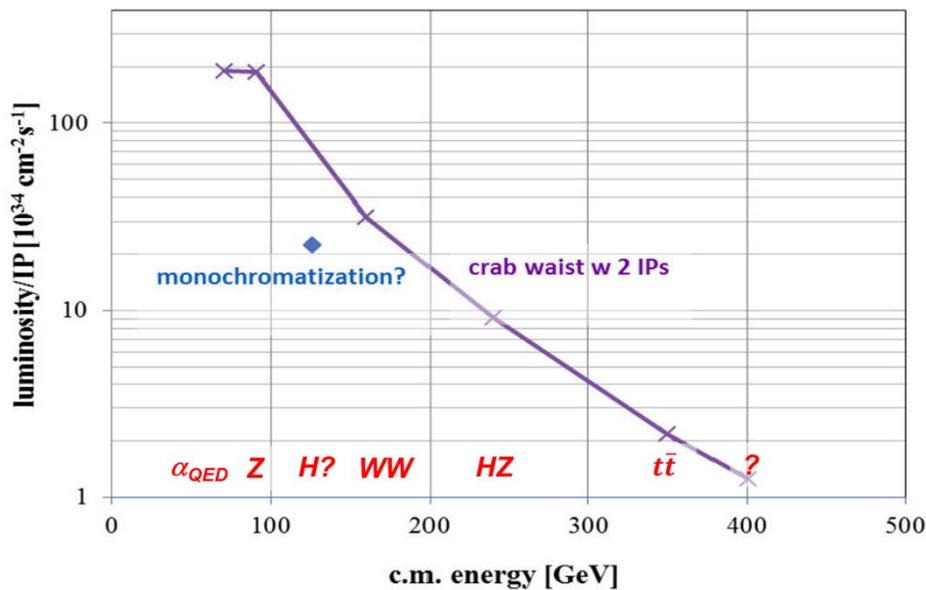

**Figure 1:** Projected FCC-ee luminosity per interaction point (IP) as a function of centre-of-mass energy, for a scenario with crab-waist collisions at two IPs.

One distinct feature of the FCC-ee design is its conception as a double ring, with separate beam pipes for the two counter-rotating (electron and positron) beams, resembling, in this aspect, the high-luminosity B factories PEP-II, KEKB and SuperKEKB as well as the LHC. The two separate rings do not only permit operation with a large number of bunches, up to a few 10,000's at the $Z$ pole, but also allow for a well-centered orbit all around the ring as well as for a nearly perfect mitigation of the energy sawtooth, e.g. by tapering the strength of all magnets according to the local beam energy, and for an independent optics control for the two beams. A side benefit at low energies is a reduction of the machine impedance by a factor of twos.

A long list of optics and beam dynamics challenges for FCC-ee includes the following: (1) final focus optics design with a target vertical IP beta function of 1 or 2 mm, 50 or 25 times smaller than for LEP2, incorporating sextupoles for crab-waist; (2) synchrotron radiation in the final focus systems and the arcs, with effects on the detector (background, component lifetime) and on the beam (vertical emittance blow up and dynamic aperture); (3) beam-beam effects, including single-turn and multi-turn beamstrahlung; (4) design of the interaction region with a strong detector solenoid with possible compensation solenoids, a large crossing angle and a pair of final-focusing quadrupoles; (5) compatibility of the layout with the design of the hadron collider sharing the same tunnel; (6) RF acceleration system for high voltage (ZH, tt) and high current (Z, WW) with possible staging scenario; (7) impedance, HOM losses and instabilities, especially for high-current "low-energy" operation at the $Z$ pole; (8) the top-up injection scheme; (9) achieving the dynamic aperture required for adequate beam lifetime and for the top-up injection, comprising the optimization of the arc optics; (10) vertical emittance control, including alignment and field errors, lattice nonlinearities, as well as beam-beam effects; (11) energy calibration and transverse polarization; (12) adapting to a non-planar tunnel; and (13) the development of a mono-chromatization for direct $H$ production in the $s$ channel. In the following we consider some of these challenges.

**Collider Layout**

Figure 2 presents one possible FCC-ee collider layout, with two collision points. The latter are located at diametrically opposed positions of the ring. The incoming beam line is less bent than the outgoing beam line in order to minimize the synchrotron radiation emitted in the direction of the experimental detector. This leads to a rather large separation of the inner and outer beam lines on each side of each interaction point (IP), most likely necessitating two separate tunnels over a distance of 5-6 km around each IP. The outer tunnel might accommodate the detector-bypass for the booster ring, as sketched in the figure, and it might later host the hadron collider. The outer and inner beam lines cross in the long straight sections half way between the two experiments. This provides a perfect two-fold symmetry of the FCC-ee collider ring, with a correspondingly decreased number of systematic resonances.

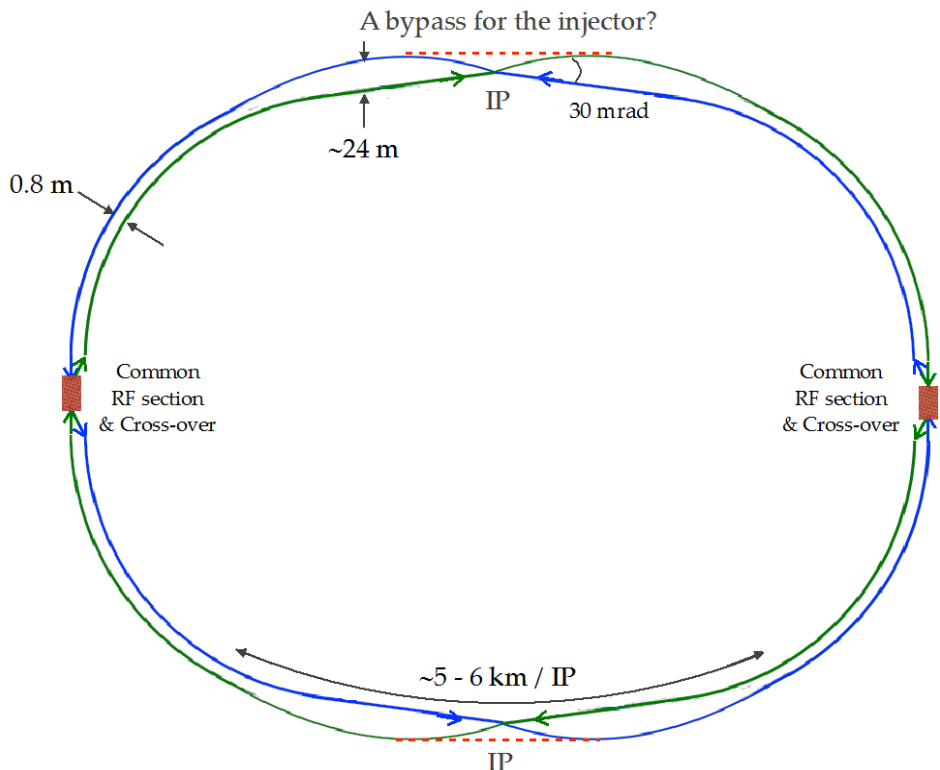

**Figure 2:** One possible FCC-ee layout (K. Oide).

**SC RF System**

The superconducting RF system is the key technology of the FCC-ee [7]. The RF system requirements are characterized by two regimes – (1) high gradients for $H$ and $t\bar{t}$ up to ≈11 GV when operating with a few tens of bunches, and (2) high beam loading with currents of about 1.5 A at the Z pole. The project aims at SC RF cavities with gradients of ≈20 MV/m, but lower gradients (e.g. 10-15 MV/m) are also acceptable. An RF frequency of 400 MHz has been chosen, equal to the one of the FCC-hh hadron collider.

The conversion efficiency from wall plug to RF power is an important figure for the overall power consumption of the facility. The FCC R&D target is 75% or higher. An efficiency of 65% was achieved for LEP2. Recent innovations in klystron design may allow for much higher values still [8].

**Staging**

Staging scenarios are being considered, in which the RF system is varied in steps, starting at low energy, e.g. Z pole operation (45.5 GeV/ beam), with fewer cavities (and correspondingly lower impedance), installing the full 400-MHz RF system for *ZH* running (120 GeV/beam), and later, for $t\bar{t}$ operation (175 GeV per beam) sharing the RF cavities for both beams, as indicated in Fig. 2, or adding higher harmonic 800 MHz cavities [9,10]. Complementary staging possibilities exist for the arc optics (varying cell length, or emittance) and for the vertical IP beta function, $\beta_y^*$.

**Final-Focus Optics**

Various final-focus optics for FCC-ee have been developed and evaluated [11,12]. A recent design is illustrated in Fig. 3, which shows the incoming half of one possible final focus optics, corresponding to the layout of Fig. 1 at a total crossing angle of 30 mrad. As indicated at the bottom, the critical photon energies for this design are below 100 keV over the last 900 m before the IP, and less than 1 kW of synchrotron radiation power is emitted here, so that all design requirements inferred from LEP experience [13,14] appear to be met. The crab waist collision scheme can be realized by a dedicated crab-waist sextupole [10] or by a "virtual" crab-waist sextupole as in Fig. 3 (based on the odd-dispersion scheme for chromaticity correction [15], where only one of two vertical chromatic correction sextupoles is located at a place with nonzero dispersion).

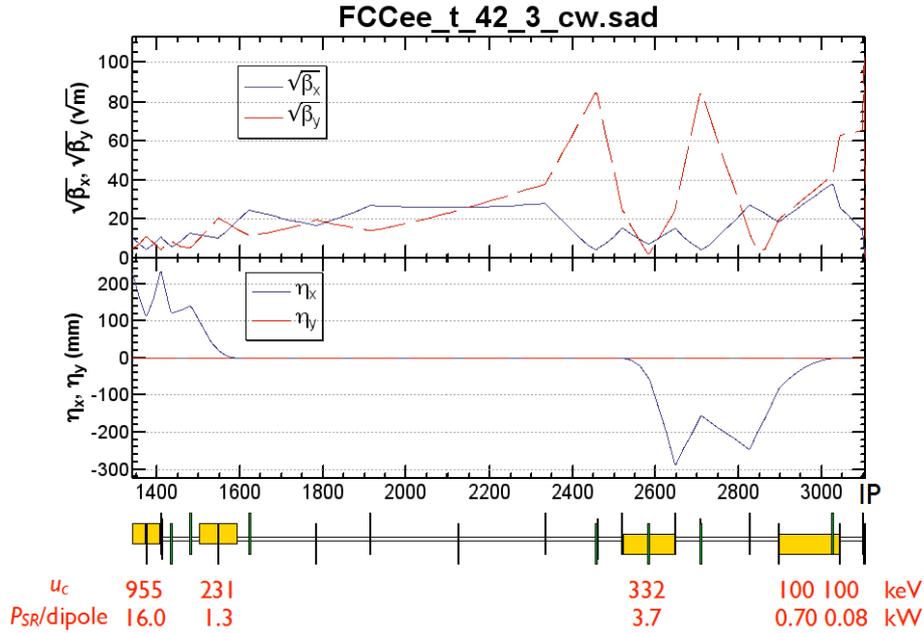

**Figure 3:** Incoming FCC-ee IR optics with low synchrotron radiation (K. Oide).

**Interaction Region**

The part of the interaction region closest to the IP is particularly challenging, due to the combination of a small $\beta_y^*$ of 1-2 mm and a large crossing angle of 30 mrad, which enhances the effect of fringe fields, kinematic nonlinearities, and synchrotron radiation. Additional complications arise from the detector solenoid field, and the need for shielding solenoids (around the final quadrupoles) as well as for an anti-solenoid (to compensate the solenoid-induced betatron coupling), together with synchrotron radiation emitted in these elements, and especially in their fringe fields [12]. Figure 4 shows one proposed configuration.

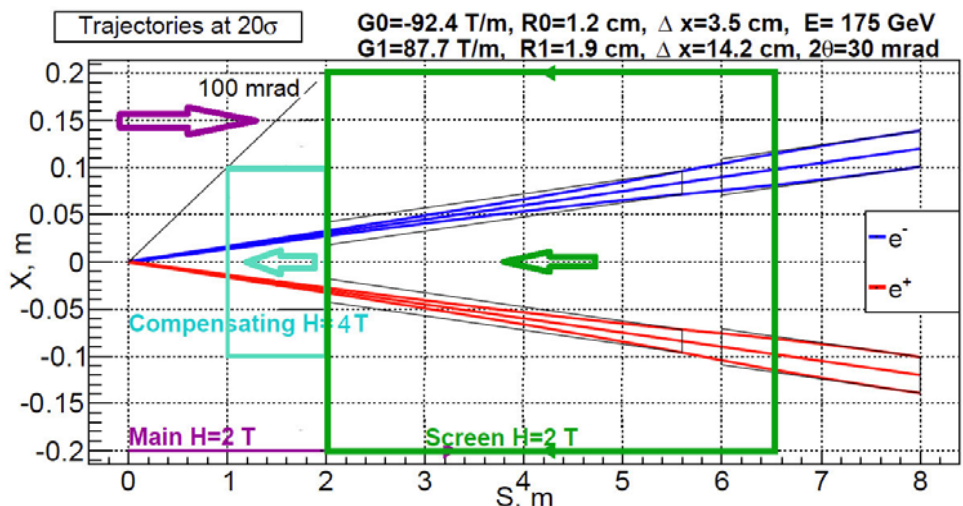

**Figure 4:** Example IR layout including main, compensating and screening solenoids (A. Bogomyagkov, S. Sinyatkin).

**Machine Detector Interface and IR Synchrotron Radiation**

Tools based on GEANT have been developed to model the machine detector interface and beam-related detector background in FCC-ee [13,14]. LEP Experiences call for critical photon energies below 100 keV and total power levels below 1 kW emitted in the direction of the particle-physics detector.

**Dynamic Aperture**

Off-momentum dynamic aperture is an important design constraint. A large acceptance improves the beam lifetime at the top threshold where beamstrahlung is important [16], and also provides space for off-momentum (top-up) injection.

The minimum required momentum acceptance, in view of beamstrahlung, is ±1.5% at 175 GeV and ±1.0% at 120 GeV, for the presently assumed beam parameters in case of crab-waist collisions.

Over the past years, the off- and on-momentum dynamic aperture of several alternative collider optics have been steadily improved, e.g. by optimizing the arc-cell phase advance and by adjusting the strengths of the arc sextupoles.

Synchrotron-radiation damping must be taken into account when simulating the dynamic aperture. Figure 5 shows an example result.

The dynamic aperture and the dynamic energy acceptance are almost acceptable in the latest optics designs. The radiation damping plays an important role for the dynamic aperture. The quadrupole fringe fields and kinematic terms can be compensated by two IR octupoles. The dynamic aperture is limited by the combined effect of IR sextupoles and arc sextupoles.

A potential issue is the energy sawtooth due to synchrotron radiation, varying from a negligible value to about 2% per half-turn from the $Z$ energy to the $t\bar{t}$ beam energy. In the two-ring scenario based on separated magnetic systems, this effect can be mitigated by varying the magnet strengths according to the local beam energy. Detailed studies of possible powering schemes are required to ensure that the momentum aperture remains sufficiently large. If during $t\bar{t}$ running the RF sections are combined for both beams, the two optics in the common regions can be matched simultaneously, as is routinely done for energy-recovery linacs.

Synchrotron radiation in the quadrupole magnets is another important effect for FCC-ee as it already was for LEP2 [17]. Indeed, this effect sets a minimum length for the arc quadrupoles. For large-amplitude particles it also leads to a breakdown of the geometric and chromo-geometric cancellations between paired sextupoles.

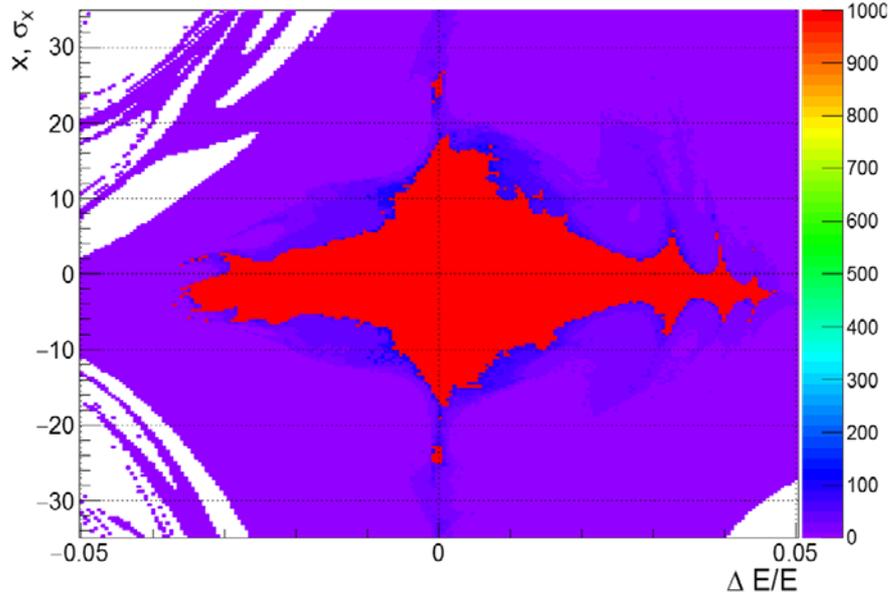

**Figure 5:** Simulated horizontal dynamic aperture as a function of initial relative momentum offset, ranging from -5% to +5%, for one FCC-ee candidate optics at 175 GeV beam energy, obtained by tracking over 1000 turns, including synchrotron motion, radiation damping, and crab-waist sextupoles. The color code indicates the number of turns survived (P. Piminov, A. Bogomyagkov).

**Beam-Beam Effects**

The crab-waist collision scheme is predicted to increase the maximum value of the vertical beam-beam tune shift at which the vertical beam size starts to blow up by about a factor of two, as compared with a standard (head-on) collision scheme.

A novel phenomenon for circular colliders is beamstrahlung, which at high energies affects the beam lifetime [16], and at low beam energies increases the bunch length and the energy spread [18,19]. Both effects are taken into account in the FCC-ee design optimization.

According to LEP experience and confirmed by some simulations, the beam-beam limit for classical head-on collisions increases with beam energy or damping decrement [20]. For FCC-ee crab-waist collisions, reducing the number of IPs from 4 to 2 may increase the maximum tune shift per IP only by a moderate 5-10% and the corresponding luminosity per IP by a similar factor [21].

**Top-Up Injection**

Top-up injection is an integral part of any high-luminosity circular collider [2]. Longitudinal injection can profit from faster damping and may have less impact on the particle-physics detector (since the design dispersion at the collision point is zero). Longitudinal injection has successfully been used at LEP [22,23]. Initial design considerations for the FCC-ee longitudinal injection include multipole kicker injection and septum-less injection schemes [24]. An alternative vertical injection scheme could potentially take advantage of the extremely small vertical emittance.

**Mono-Chromatization**

An interesting options presently under study is the possibility of direct Higgs production in the *s* channel, at a beam energy of 63 GeV. In order to obtain an acceptable Higgs event rate and to precisely measure the width of this particle mono-chromatization will be required. The mono-chromatization can be realized, e.g., by introducing horizontal IP dispersion of opposite sign for the two colliding beams [25,26]. The mono-chromatization factor should be larger than 10.

**Impedance and Instabilities**

Impedance effects are a concern, in particular for the high-current operation at the *Z* pole. The energy loss at the RF cavities can be as large as the energy loss due to synchrotron radiation [27]. Fortunately, most of the power will be dissipated in the tapers outside the low-temperature cavity cells. Higher-order mode (HOM) heating of the cavities is a related concern, calling for efficient HOM dampers operating at room temperature. As this has the potential to limit the beam current—thus the maximum luminosity achievable at the *Z* pole—we will continue to investigate means to reduce the loss factor.

In addition the heavy-beam loading and residual HOM-driven instabilities require strong longitudinal feedback loops, perhaps similar to those for PEP-II, while a transverse bunch-by-bunch feedback must suppress resistive-wall, HOM-driven, and ion instabilities. Both the B factories as well as the LHC have demonstrated transverse damping times on the order of 10 turns, which gives a measure of the maximum undamped growth rate allowable.

At LEP the transverse mode coupling instability at injection limited the achievable bunch intensity. By contrast, at FCC-ee the beam is always at full collision energy.

**Polarization and Energy Calibration**

Scaling from LEP some natural transverse polarization due to the Sokolov-Ternov effect is expected up to the *W* threshold (80 GeV / beam) or above. In this energy range resonant depolarization of a few dedicated non-colliding bunches will provide an exquisite measurement of the average beam energy [28]. Extrapolation to the beam energy at the IPs, taking into account the energy sawtooth as well as possible beam-beam effects, may lead to some systematic uncertainties. For higher beam energy and as a cross-check other techniques, such as Compton backscattering schemes and also measuring the spin precession of an injected polarized beam [29], are being considered. These techniques would also allow for a cross calibration.

The potentially harmful effect of an orbit kink on the polarization and on the vertical emittance can be avoided by a special orbit inclination technology [30]: Twists between arc segments match the horizontal plane of oscillations with the bending planes of the segments. Spin matching is provided by weak solenoids which produce roughly half of the full twist. The other half of the twist is obtained from a unity/minus-unity insertion appropriately rotated around the longitudinal axis [30].

**Super KEKB Test Bed**

SuperKEKB [31,32] will be an important demonstrator for a number of key concepts of the *FCC-ee* design. Simply speaking, all elements not yet tested at LEP2, KEKB or PEP-II will be demonstrated by SuperKEK.

In various regards SuperKEKB actually goes beyond FCC-ee. For example, SuperKEKB will implement top-up injection at higher current with a shorter beam lifetime. The $\beta_y^*$ of SuperKEKB will be 300 μm, to be compared with 1 or 2 mm at FCC-ee (see Fig. 6). The design beam lifetime is 5 minutes, limited by Touschek scattering, while the *FCC-ee* beam lifetime is more than 20 minutes, due to radiative Bhabha scattering (and to some extent beamstrahlung). SuperKEKB aims at a vertical-to-horizontal emittance ratio of 0.25% with colliding beams, similar to FCC-ee. The off-momentum design acceptance of SuperKEKB is ±1.5%. Such a value would also be sufficient for FCC-ee operation at the $t\bar{t}$ threshold, where beamstrahlung may have a noticeable effect on the beam lifetime [16]. The SuperKEKB-injector e$^+$ production rate of $2.5\times10^{12}$/s is even higher than required for *FCC-ee* crab-waist running on the Z pole ($<1.5\times10^{12}$/s). The SuperKEKB beam commissioning will start in early 2016.

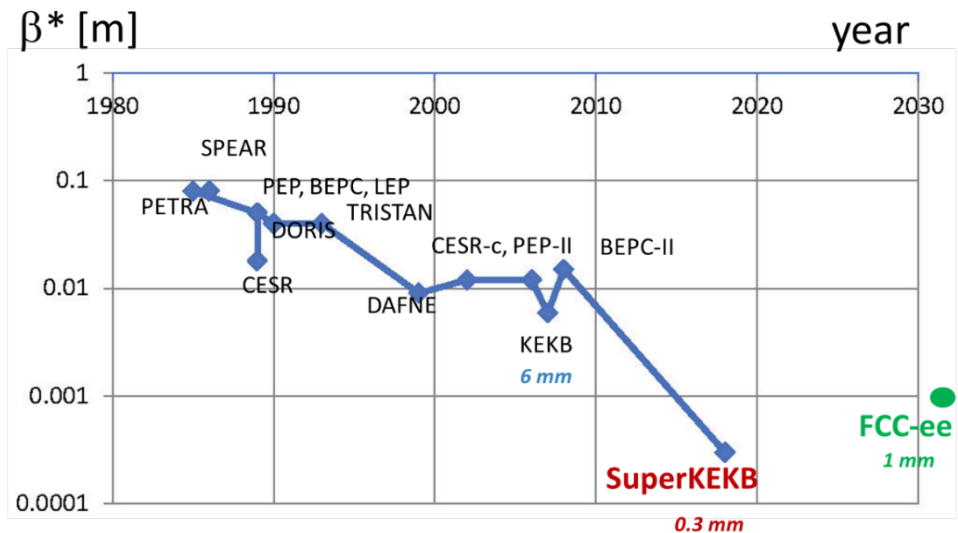

**Figure 6:** $\beta_y^*$ evolution in circular e$^+$e$^-$ colliders over 50 years, including the upcoming SuperKEKB and FCC-ee.

**Conclusions and Outlook**

Over the past years the optics development and beam dynamics studies for FCC-ee have made great progress. A double ring collider with crab waist collisions promises superb performance over a large range of beam energies, and allows for an elegant staging.

The primary design challenges arise from the tight focusing, the large energy acceptance required, the wide range of beam parameters and beam energies to be accommodated, severe constraints on the final-focus synchrotron radiation, the effects of the detector solenoids and their compensation, polarization issues, and the required compatibility with the layout of the FCC-hh hadron collider.

**Acknowledgements**

This work was supported, in parts, by the European Commission under the Capacities 7th Framework Programme project EuCARD-2, grant agreement 312453, and the HORIZON 2020 project EuroCirCol, grant agreement 654305.

**References**

1. Sixteenth European Strategy Session of Council, Brussels, 30 May 2013, CERN-Council-S/107/Rev. (2013)
2. A. Blondel, F. Zimmermann, "A High. Luminosity e+e- Collider in the LHC tunnel to study the Higgs Boson," CERN-OPEN-2011-047, arXiv:1112.2518v1 (2011).
3. J. Wenninger et al., "Future Circular Collider Study — Lepton Collider Parameters," FCC-ACC-SPC-0003 rev. 2.0 (2014).
4. A. Bogomyagkov, E. Levichev, and D. Shatilov, "Beam-Beam Effects Investigation and Parameter Optimization for Circular e+e− Collider TLEP to study the Higgs Boson," Phys. Rev. ST Accel. Beams 17, 041004 (2014).
5. A. Blondel, P.Janot, et al., see, e.g., FCC Week 2015, Washington DC, 23-29 March 2015, http://indico.cern.ch/event/340703/
6. U. Wienands, "Is polarization possible at TLEP?," 4th TLEP workshop, CERN, 4-5 April 2013, http://indico.cern.ch/event/240814/timetable/#20130405
7. A. Butterworth,"The RF System for FCC-ee," EPS-HEP 2015 Vienna, Jul 2015
8. I. Syratchev, "High Efficiency Klystron Development," High Gradient workshop, Beijing, June 2015
9. U. Wienands, "Staging Scenarios for FCC-ee," at Aspen Winter Conference "Exploring the Physics Frontier with Circular Colliders," 31 January 2015.
10. M. Benedikt et al., "Combined Operation and Staging for the FCC-ee Lepton Collider," Proc. IPAC'15 Richmond (2015)
11. R. Martin et al. "Status of the FCC-ee Interaction Region Design," Proc. IPAC'15.
12. A. Bogomyagkov et al., Interaction Region of the Future Electron-Positron Collider FCC-ee," Proc. IPAC'15 Richmond.
13. H. Burkhardt, "Synchrotron Radiation in Interaction Region + Comment on BBREM (Rad. Bhabha) Lifetime," FDCC Week 2015, Washington http://indico.cern.ch/event/340703/session/25/contribution/15#preview:919299
14. H. Burkhardt, M. Boscolo, "Tools for Flexible Optimisation of IR Designs with

Application to FCC," Proc. IPAC'15 Richmond.
15. K. Oide, "Final Focus System with Odd-Dispersion Scheme," Proc. 5th International Conference on High-Energy Accelerators, Hamburg, Germany, 20 - 24 Jul 1992, pp.861-863 and KEK-92-58 (1992).
16. V. Telnov, "Restriction on the energy and Luminosity of $e^+e^-$ Storage Rings due to Beamstrahlung," PRL 110, 114801 (2013).
17. F. Barbarin, F.C. Iselin, J.M. Jowett, "Particle Dynamics in LEP at Very High Energy," Proc. 4th European Particle Accelerator Conference, London, UK, 27 Jun - 1 Jul 1994, pp.193-195
18. K. Yokoya, "Scaling of High-Energy e+e- Ring Colliders," KEK Accelerator Seminar, 15 March 2012.
19. K. Ohmi, F. Zimmermann, "FCC-ee/CepC Beam-Beam Simulations with Beamstrahlung," Proc. IPAC2014, p. 3766.
20. R. Assmann, K. Cornelis, "The Beam-Beam Interaction in the Presence of Strong Radiation Damping," Proc. EPAC2000 Vienna, p. 1187.
21. D. Shatilov, "Beam-Beam Effects in High-Energy Colliders: Crab Waist vs. Head-On," Proc. ICFA HF2014, 9–12 October 2014, Beijing (2014); and private communication (2015).
22. P. Collier, Synchrotron Phase Space Injection into LEP, Proc. 16th Biennial Particle Accelerator Conference and International Conference on High-Energy Accelerators, Dallas, TX, USA, 1 - 5 May 1995, pp.551-553
23. P. Baudrenghien, P. Collier, "Double Batch Injection into LEP," Proc. 5th European Particle Accelerator Conference, Sitges, Barcelona, Spain, 10 - 14 Jun 1996, pp.415-417
24. M. Aiba et al., "Top-Up Injection for FCC-ee," CERN-ACC-2015-065 (2015).
25. J. Jowett, "Feasibility of a Monochromator Scheme in LEP," LEP Note 544 (1985).
26. A. Faus-Golfe, J. Le Duff, "Versatile DBA and TBA lattices for a Tau-Charm Factory with and without Beam Monochromatization, Nuclear Inst. and Methods in Physics Research, A, Volume 372, Issue 1 (1996) p. 6-18
27. M. Migliorati, "Impedance and Collective Effects," FCC Week 2015, Washington http://indico.cern.ch/event/340703/session/81/contribution/158/material/slides/0.pdf
28. U. Wienands, "Is polarization possible at TLEP?," 4th TLEP workshop, CERN, 4-5 April 2013, http://indico.cern.ch/event/240814/timetable/#20130405
29. I.A. Koop, "Scenario for Precision Beam Energy Calibration in FCC-ee," CERN-ACC-2015-081 (2015).
30. I.A. Koop, "Mitigating the Effect of an Orbit Kink on Vertical Emittance and Polarization," CERN-ACC-2015-090 (2015).
31. T. Miura et al., "Progress of SuperKEKB," Proc. IPAC'15 Richmond (2015).
32. SuperKEKB web site: http://www-superkekb.kek.jp/ .